\documentclass[prb,aps,showpacs,eqsecnum]{revtex4}

\usepackage{graphicx}
\usepackage{dcolumn}
\usepackage{bm}

\begin{document}
\title{Memory effects in superfluid vortex dynamics}

\author{H. M. Cataldo}
\email[Corresponding author. Electronic address: ]{cataldo@df.uba.ar}

\author{D. M. Jezek}

\affiliation{Departamento de F\'{\i}sica, Facultad de Ciencias Exactas y Naturales, \\
Universidad de Buenos Aires, RA-1428 Buenos Aires, Argentina\\
and Consejo Nacional de Investigaciones Cient\'{\i}ficas y
T\'ecnicas, Argentina}

\date{\today}

\begin{abstract}
The dissipative dynamics of a vortex line in a superfluid is investigated within the frame of
a non-Markovian quantal Brownian motion model. Our starting point is a recently proposed 
interaction Hamiltonian between the vortex and the superfluid quasiparticle excitations, 
which is generalized to incorporate the effect of scattering from fermion impurities ($^3$He
atoms). Thus, a non-Markovian equation of motion for the mean value of the vortex position
operator is derived within a weak-coupling approximation. Such an equation is shown to yield,
in the Markovian and elastic scattering limits, a $^3$He contribution to the longitudinal
friction coefficient equivalent to that arising from the Rayfield-Reif formula.
Simultaneous
Markov and elastic scattering limits are found, however, to be incompatible, since an 
unexpected breakdown of the Markovian approximation is detected at low cyclotron frequencies.
Then, a non-Markovian expression for the longitudinal friction coefficient is derived and
computed as a function of temperature and $^3$He concentration. Such calculations show that
cyclotron frequencies within the range 0.01$-$0.03 ps$^{-1}$ yield a very good agreement to the
longitudinal friction figures computed from the Iordanskii and Rayfield-Reif formulas for pure
$^4$He, up to temperatures near 1 K. A similar performance is found for nonvanishing $^3$He
concentrations, where the comparison is also shown to be very favorable with respect to the
available experimental data. Memory effects are shown to be weak and increasing 
with temperature
and concentration.
\end{abstract}
\maketitle

\section{Introduction}
\label{sec1}
It is widely accepted that the superfluid vortex dynamics at zero temperature
is ruled by the Magnus force:\cite{don}
\begin{equation}
\label{1.1}
m_v\dot{\bf v}=\rho_s\kappa\hat{\bf z}\times({\bf v}-{\bf v}_s).
\end{equation}
Here we are assuming a straight vortex line parallel to the $z$ axis, moving with a velocity
${\bf v}$. ${\bf v}_s$ denotes a uniform background superfluid velocity which, in the simplest
case, may be assumed to be time-independent and then dropped from Eq. (\ref{1.1}), if ${\bf v}$
is reinterpreted by Galilean invariance as a vortex velocity relative to a background 
superfluid at rest. $\rho_s$ denotes the superfluid mass density and $\kappa$ the 
quantized circulation of the vortex 
velocity field (e.g., $\kappa=h/m_4$ for one quantum of counterclockwise 
circulation, being $m_4$ the mass of a $^4$He atom and $h$ the Planck's constant).
Then, the right-hand side of Eq. (\ref{1.1}) represents the Magnus force per unit length 
acting on
the vortex and, accordingly, $m_v$ on the left-hand side represents an effective vortex 
mass per 
unit length. However, there is no consensus in the literature as regards the value of $m_v$.
Most of the treatments so far,
have neglected $m_v$ by assuming that it must be equivalent to the 
hydrodynamic mass of a core of atomic dimensions.\cite{don}
Then, from (\ref{1.1}) we get the well-known law of zero temperature
vortex dynamics, which states
that a vortex must move at the velocity which the superfluid possesses at the location of the
vortex itself.\cite{noz}
 On the other hand, more recent theories\cite{duan,arovas,tang} claim that the
vortex mass should not be ignored, since it is shown 
to be logarithmically divergent with the system
size, thus exceeding by far the core mass. Such a large mass, however, can be shown to be 
consistent
with the above law of vortex dynamics, if the dissipative mechanisms
acting at zero temperature are taken into
account. In fact, one must consider the effect of 
the vortex coupling to the
superfluid which should give rise to dissipation in the form of
phonon emission, in close analogy to the photon radiation
mechanism stemming from an accelerated charge in electrodynamics.\cite{arovas,vinen} 
Another dissipative mechanism should arise in ordinary helium from the vortex coupling to the
Fermi gas of $^3$He impurities. Whatever the case, it is remarkable that even for an unbounded
system leading to a divergent vortex mass, such dissipative mechanisms should make the vortex
reach the superfluid velocity, in accordance with the fundamental law of zero temperature
vortex dynamics (see Sec. \ref{VC}).

At nonvanishing temperatures, in addition to phonon radiation and $^3$He scattering, there
exists
 a third dissipative mechanism stemming from the vortex scattering of superfluid quasiparticle
 excitations (phonons and/or rotons).

Now, let us return to the Eq. (\ref{1.1}) and note that for a background superfluid at rest
(${\bf v}_s=0$), the vortex dynamics turns out to be identical to the two dimensional one
of an electron moving in a uniform magnetic field subjected to the Lorentz force.
Then, expressing the two dimensional vector ${\bf v}$ in complex notation as
$V=v_x+iv_y$, we can rewrite (\ref{1.1}) in the more compact form
\begin{equation}
\label{1.2}
\dot{V}=i\Omega V,
\end{equation}
which clearly shows that the vortex will move in a circle at the angular 
frequency (cyclotron frequency)
\begin{equation}
\Omega=\frac{\rho_s\kappa}{m_v}.
\label{1.3}
\end{equation}
The above mechanisms of dissipation lead to a complex shift of $\Omega$, according to which
(\ref{1.2}) becomes
\begin{equation}
\label{1.4}
\dot{V}=(i\Omega_{\rm eff}-\nu_d) V,
\end{equation}
where $\Omega_{\rm eff}$ denotes the effective angular frequency into which the unperturbed
cyclotron frequency is shifted, and $\nu_d>0$ represents a damping frequency that sets the
time scale at which $V$ tends to zero. The above equation, however, can be written in a more
familiar form if we return to the vector notation of Eq. (\ref{1.1}):
\begin{equation}
\label{1.4p}
m_v\dot{\bf v}=(\rho_s\kappa-D')\hat{\bf z}\times{\bf v}-D{\bf v},
\end{equation}
where
\begin{eqnarray}
\label{1.4p1}
D'&=& \rho_s\kappa(1-\Omega_{\rm eff}/\Omega) \\
\label{1.4p2}
D &=& \rho_s\kappa\nu_d/\Omega
\end{eqnarray}
respectively denote {\it transverse} and {\it longitudinal}
 friction coefficients,\cite{don,bar} and we have assumed
 that the normal fluid remains at rest.
Actually, a vortex in motion may drag the normal fluid in its vicinity, but this effect
should be negligible below 1 K and we shall restrict our study to such situation.\cite{bar}

At this point, it is important to notice that the
cyclotron motion represented by Eq. (\ref{1.2}) is also characteristic of the helical waves
on vortex lines and rings, usually known as Kelvin waves.\cite{don,flu,prog} In fact, each 
vortex line element in such waves executes motion about the undisturbed line in a circle of 
radius $d$ and with a frequency $\omega$, which approximately fulfil
\begin{equation}
\label{1.5a}
m_v\omega^2d=\rho_s\kappa v_i+\rho_s\kappa \omega d,
\end{equation}
where the amplitude of the deformation $d$ is assumed to be much less than the wavelength 
$\lambda$. The above equation corresponds to the centripetal component of an expression of the
form (\ref{1.1}), where now ${\bf v}_s=-v_i\,{\bf\hat{\theta}}$ denotes the local self-induced 
velocity,\cite{hama} which points in a direction opposite to the one of the superfluid 
velocity field generated by the undisturbed vortex line. Therefore, the line velocity in 
(\ref{1.1}) may be expressed as ${\bf v}=\omega d\,{\bf \hat{\theta}}$,
 where the frequency $\omega$
will be nonpositive if ${\bf v}$ points in the same sense as the self-induced velocity.
In fact, the solution of the quadratic equation (\ref{1.5a}) yields two frequency branches of
opposite sign:
\begin{equation}
\label{1.5b}
\omega_\pm=\frac{\rho_s\kappa}{2m_v}\left[1\pm\sqrt{1+\frac{4v_im_v}{
\rho_s\kappa d}}\right],
\end{equation}
whose physical meaning can be easily understood in the limit of long wavelengths ($
v_im_v/(\rho_s\kappa d)<<1$, $v_i\sim \kappa d/\lambda^2$). That is, the positive fast branch
$\omega_+\simeq \Omega$ corresponds to the cyclotron motion previously described, while the 
negative slow branch $\omega_-\simeq -v_i/d\sim -\kappa/\lambda^2$ corresponds to the motion
of the vortex element with its local self-induced velocity. Then we may see that for a 
massless vortex line only the slow
branch would exist, 
this being the common assumption in the literature of helium vortex waves.
As regards experimental studies, only the slow branch has been detected by means of a resonant
coupling to transverse radio-frequency electric fields acting on vortex lines charged with
ions.\cite{ashton} On the other hand, the thermal excitation of Kelvin waves has been 
theoretically investigated with rather surprising results.\cite{flu}
In fact, it was found that the 
entropy of such waves increases above temperatures about 1.85 K, so that the 
free energy of the vortices is driven negative, with the consequence that superfluidity would
be destroyed. This phenomenon has been called the ``free energy catastrophe'' and the authors
suggest that it could arise from their
 neglect of the effect of the vortex line on the neighboring
phonons and rotons in the system. Now, given that the major contribution to the 
free energy comes from the slow branch, such a ``catastrophe'' would apparently be shifted
towards temperatures above the lambda transition if only the excitation of the cyclotron
branch were taken into account. Actually, this has implicitly
 been assumed by most of the studies
on thermal excitations of vortices through phonon and roton scattering, since the authors 
have only considered straight vortex lines. In particular, the phonon scattering excitation
of the slow branch was analyzed by Fetter\cite{fetter} and
 Sonin,\cite{son} who concluded that it yields only a small
correction to the friction force calculated for a rectilinear vortex. In fact, the former
restriction to considering only straight vortex lines in calculations of the friction 
coefficients, arises naturally if we accept the {\em basic premise} that it is only the
relative motion of an {\em element} of line with respect to the normal fluid what matters in
such type of calculations.
 That is, any relative motion of a vortex line element should be 
subjected to the same kind of friction force per unit length, i.e., the same value of the
friction coefficients $D$ and $D'$. Following these considerations, we have focused our 
calculations on the simplest situation of a damped cyclotron motion of a straight vortex line.

There are, to our knowledge, no experimental results on the transverse friction 
coefficient below 1.3 K. On the other hand, as regards the longitudinal friction coefficient,
we must refer to the pioneering experiments 
performed by Rayfield and Reif (R-R) 
in the early sixties.\cite{reif} In fact, they studied the temperature
dependence of the rate of energy loss of charge-carrying vortex rings moving through 
helium II. The radii of such rings are large ($>500$ \AA) compared to the distance over
which a vortex line is expected to interact appreciably with a quasiparticle. Hence the 
frictional
forces on these vortex rings must be the same as those on vortex lines bent into circles.
So, R-R were able to measure what they called the {\it attenuation coefficient} $\alpha$,
which turns out to be
 simply proportional to the longitudinal friction coefficient ($\alpha=\kappa
D/2$). Here it is important to notice that the energy losses in the R-R experiment 
are consistent with
a friction owing to the axial displacement of rings, i.e., the main relative motion of each
line element with respect to the normal fluid will not correspond to the cyclotron
motion. We shall see, however, that in accordance with the above basic premise, the 
longitudinal
friction coefficient arising from our theory shows an excellent agreement with the one 
arising from the R-R attenuation coefficient $\alpha$.
We notice also that for a cyclotron motion, the radiation damping should be at least comparable
to the scattering one for temperatures below 1 K (see Sec. \ref{VC}), even though we shall 
focus exclusively on the scattering processes, since phonon emission is supposed to
 be negligible
for the axial displacements in the R-R experiment.

Using kinetic-theory arguments, R-R showed that $\alpha$ comprises three terms, one
for each class of
quasiparticle interacting with the vortex, namely phonons, rotons and $^3$He impurities.
Each of such contributions was shown to be proportional to a corresponding averaged cross
section over all momenta, and R-R could determine by fitting to their experimental results,
that the roton and $^3$He cross sections are approximately temperature-independent, with
respective values 9.5 and 18.3 \AA. The R-R experiments were performed in the range of 
temperatures between 0.28 and 0.7 K, and $^3$He concentrations between 1.4$\times$10$^{-7}$
(ordinary helium) and 2.84$\times$10$^{-5}$. Then, at the lowest temperatures and highest
$^3$He concentrations, only the $^3$He contribution to $\alpha$ is appreciable, allowing
its separate study. Analogously, in the opposite limit of high temperatures and low 
$^3$He concentrations, only the roton contribution to $\alpha$ survives. Unfortunately,
only scant information as regards the phonon contribution to $\alpha$ could be derived from
such experiments, since even though phonon scattering is dominant at the lowest temperatures
in pure $^4$He, the scattering from $^3$He impurities becomes the most important contribution
in ordinary helium. R-R employed Pitaevskii's\cite{pita} calculation of the 
phonon-scattering cross section to evaluate the phonon contribution to $\alpha$, but soon
after the publication of R-R's paper, Iordanskii\cite{ior} reported  an improved theory
of the frictional force due to phonons, which seems to be so far the most reliable one.
Both, Iordanskii's theory and the above kinetic-theory analysis of R-R are based upon
an {\it elastic scattering assumption}, by which the energy of any quasiparticle that collides
 with the
vortex is conserved after the collision. This amounts to ignoring any energy the vortex could
exchange in such a process, in particular the cyclotron energy quantum $\hbar\Omega$, which
then should be negligible
 with respect to the energy
of any quasiparticle colliding with the vortex.  In conclusion, 
 one should expect a cyclotron frequency of finite value, most likely
compatible with an elastic scattering approximation. Such a hypothesis, has been recently
put forward in Ref.~\onlinecite{ca1}
 (henceforth to be designated as $I$), where we have studied the friction
arising from the scattering of superfluid quasiparticle excitations in the form of a
translationally invariant interaction potential. Then, the first order expansion in the vortex
velocity of such a potential was shown to yield vortex transitions between nearest Landau
levels, mediated by one-quasiparticle transitions.
Thus, in the frame of such a model of quantal Brownian motion for the vortex dynamics, the
longitudinal friction coefficient was computed by making use of weak-coupling and Markov
approximations. The result was shown to be equivalent, in the limit of elastic scattering,
to that arising from the Iordanskii formula and, proposing a simple functional form for
the scattering amplitude, with a single adjustable parameter 
whose value was set to get agreement to the
Iordanskii result for phonons, an excellent agreement with experimental data was found,
up to temperatures
about 1.5 K. Finite values of the cyclotron frequency of order 0.01 ps$^{-1}$ were
also shown to yield practically the same results.

In the present article, we 
pursue such an investigation in order to analyze the incidence of
vortex-$^3$He scattering, which, as mentioned for ordinary helium, turns out to be the 
most important contribution to the friction at low temperatures. But more importantly,
we report an unexpected breakdown of the Markov approximation at low cyclotron frequencies,
unnoticed in previous treatments within the elastic scattering limit.
Actually, the interaction of the vortex with the remaining degrees of freedom of helium, leads
to integrodifferential equations of motion for the vortex observables,
according to which the present
vortex motion turns out to be influenced by its whole previous history. In the Markov 
approximation, such a memory is assumed to be negligible and the vortex equations of motion
are approximated by differential equations like (\ref{1.4}).\cite{note}
 We shall show that for low enough
cyclotron frequencies, the Markov approximation fails and non-Markovian or memory effects must
be taken into account. Such effects can be of importance in diverse quantum Brownian motion 
problems\cite{haake,ford,ca4,ca5} and, particularly, in physical situations which involve
Brownian models of the dynamics of charged particles. For example, a fully non-Markovian
reformulation of the Abraham-Lorentz theory of radiation reaction in electrodynamics, has been
shown to lead to the elimination
 of ``runaway solutions" and causality violations occurring in the
original theory.\cite{f2} In transport theory, the phenomenological Drude-Lorentz result for 
the ac conductivity has been shown to be affected by important memory effects, especially away
from resonance\cite{f3} 
and, in the context of two-dimensional 
magnetotransport, the classical magnetoresistance appears as a consequence of 
memory effects which are beyond the Boltzmann-Drude approach.\cite{boby,dmitri}
It may be useful to expand on the last problem, since it corresponds just to a
classical two-dimensional Brownian motion of an electron, subjected to a uniform magnetic field
perpendicular to the plane. In fact, the electron is supposed to
move through a random array of stationary
scatterers (background impurities) with short range forces, and there are memory effects of
two types: (i) the electron may recollide with the same impurity, or (ii) its trajectory may
repeatedly pass through a space region which is free of impurities. It has recently been 
shown that backscattering processes of the type (ii)  are 
responsible, at low cyclotron frequencies,
of additional memory effects leading to unexpected features of the 
magnetoresistance.\cite{dmitri} Even though there are obviously
important differences with the vortex 
problem, it is instructive to compare with this
simpler problem, where the source of memory effects at low cyclotron frequencies has been
fully recognized.

Microscopic approaches to quantal Brownian motion also show that memory effects are often
important when the weak-coupling approximation becomes poorer.\cite{haake} This point will be
analyzed for our model in Sec. \ref{VB}.

This paper is organized as follows, in the next section, starting from a straightforward
generalization to include $^3$He of the Hamiltonian utilized in $I$, a non-Markovian equation
of motion for the vortex dynamics is derived within a weak-coupling approximation.
Next, we analyze the Markov approximation and the limit of elastic scattering,
showing that under such approximations, the longitudinal friction coefficient stemming from
$^3$He scattering, can be shown to be equivalent to that arising from the corresponding
R-R formula. In Sec. III we analyze the breakdown of the Markov approximation at low cyclotron
frequencies and develop a non-Markovian
treatment, from which expressions for the longitudinal and transverse friction coefficients
are derived. In Sec. IV we focus on the simpler case of a pure $^4$He system.
We compare in Sec. \ref{secVA} our results for the longitudinal friction
 with the Iordanskii (phonon) plus the R-R (roton)
results, finding a very good agreement within the cyclotron frequency range 0.01$-$0.03
ps$^{-1}$, up to temperatures near 1 K. In Sec. \ref{VB} we study the frequency ratio
$\Omega_{\rm eff}/\Omega$, which provides a measure of the memory introduced into
the vortex dynamics. We explain also the difficulties involved in the calculation of the 
transverse friction coefficient, due to which only its order of magnitude could be estimated.
In Sec. \ref{VC},
following the theory developed by Arovas and Freire,\cite{arovas}
 we discuss the memory effects related to phonon radiation at zero temperature.
Finally, Sec. V deals with 
dilute solutions of $^3$He in $^4$He, where we compare 
our results to the available experimental data, and extend our
study of the memory parameter $\Omega_{\rm eff}/\Omega$ in the presence of $^3$He. 

\section{vortex equation of motion, markov approximation and the limit of elastic scattering}
Our starting point is the following Hamiltonian, which arises as a straightforward
generalization of the Hamiltonian proposed in $I$, in order to take into account the presence
of $^3$He impurities:
\begin{equation}
H=H_0+H_{int},
\label{H}
\end{equation}
where
\begin{equation}
H_0=\hbar\Omega\left(a^\dagger a+\frac{1}{2}\right)+
\sum_{ {\bf k}}\,\hbar\omega_k \,\,
b_{{\bf k}}^\dagger \,  b_{ {\bf k}}+
\sum_{ {\bf q},\sigma}\,\epsilon_q \,\,
c_{{\bf q},\sigma}^\dagger \,  c_{ {\bf q},\sigma}
\label{H0},
\end{equation}
and
\begin{equation}
H_{int}=\frac{2i}{\Omega}
\sum_{ {\bf k} , {\bf q},\sigma }  \, \, \delta_{k_zq_z}
[\Lambda(k,q) b_{{\bf k}}^\dagger \,  b_{ {\bf q}}+
\Gamma(k,q) c_{{\bf k},\sigma}^\dagger \,  c_{ {\bf q},\sigma}]
({\bf k}-{\bf q})\times \hat{{\bf z}}\cdot{\bf v}.
\label{Hint}
\end{equation}
$H_0$ gives the noninteracting part of the Hamiltonian and it comprises three terms, 
the first
of which corresponds to the cyclotron motion of
 the vortex line, the second one to helium II excitations, and the last one 
to
$^3$He quasiparticles, i.e., $a^\dagger$, $b_{{\bf k}}^\dagger$, and  
$c_{{\bf q},\sigma}^\dagger$ respectively denote, a creation operator of right circular quanta,
a creation operator of helium II quasiparticle excitations of momentum $\hbar{\bf k}$ and
 frequency $\omega_k$, and a creation operator of
$^3$He quasiparticles of momentum $\hbar{\bf q}$, energy $\epsilon_q$ and
spin 1/2 projection $\sigma$. The interaction Hamiltonian $H_{int}$ arises as a 
straightforward generalization of the form given in $I$,
to include the effect of vortex-$^3$He
scattering processes. In fact, if we replace in Eq. (\ref{Hint}) the vortex velocity operator
${\bf v}$ as a linear combination of creation and annihilation operators of right circular 
quanta,\cite{ca1}
 it becomes clear that the interaction consists of vortex-quasiparticle scattering events that 
make the vortex to raise or lower one Landau level.
Then, in addition to the scattering amplitude $\Lambda(k,q)$ related to the vortex
interactions with phonons and rotons discussed in $I$, now we are including a scattering 
amplitude
$\Gamma(k,q)$, which takes into account vortex-$^3$He interactions.

In previous works\cite{ca2,ca3,ca6} 
we derived, by means of a standard reduction-projection procedure
and a weak-coupling Markov approximation, a generalized master equation for the density 
operator of the vortex. Our aim was to obtain an equation of motion for the mean value of
the complex vortex position operator $R=x+iy$. Now we are interested in rederiving such an
equation of motion from
 the more general Hamiltonian (\ref{H})-(\ref{Hint}). This time 
we have employed a simpler and more direct procedure (see the Appendix), which leads to 
the following 
integrodifferential equation of motion for $v(t)\equiv
e^{-i\Omega t}\langle\dot{R}(t)\rangle$:
\begin{equation}
\dot{v}(t)+\int_0^t d\tau\,{\cal D}(\tau)v(t-\tau)=0,
\label{noM}
\end{equation}
where
\begin{eqnarray}
{\cal D}(\tau) & = &
\frac{1}{\hbar^2\pi\Omega\rho_sL/m_4}\sum_{ {\bf k}, {\bf q}}  \, \, \delta_{k_zq_z}
({\bf k}-{\bf q})^2[|\Lambda(k,q)|^2(\omega_k-\omega_q)(n_q-n_k)
e^{i(\omega_k-\omega_q-\Omega)\tau}\nonumber\\
&&+\frac{2}{\hbar}|\Gamma(k,q)|^2(\epsilon_k-\epsilon_q)(f_q-f_k)
e^{i(\epsilon_k/\hbar-\epsilon_q/\hbar-\Omega)\tau}],
\label{16}
\end{eqnarray}
being $n_k=[\exp(\hbar\omega_k/k_BT)-1]^{-1}$ and $f_k=\{\exp[(
\epsilon_k-\mu)/k_BT]+1\}^{-1}$ the thermal equilibrium Bose and Fermi
occupation numbers for the corresponding quasiparticle excitations, respectively. 

The dynamics behind Eq. (\ref{noM}) can be understood by noting that the present vortex motion
is actually influenced by its whole previous history, each time being weighted by a memory
kernel ${\cal D}(\tau)$ such that $\tau=0$ weights the present time, while $\tau=t$
weights  the initial condition. Then, if ${\cal D}(\tau)$ possesses
a microscopic lifetime $\tau_m$ compared to
the characteristic times that rule the motion of $v(t)$, only the present time $t$ will have
a nonnegligible influence upon the vortex motion, provided $t>>\tau_m$.
This constitutes the so-called {\it Markov} or 
{\it long time limit} approximation,\cite{note,haake,ford,ca4,ca5}
under which Eq. (\ref{noM}) becomes a differential equation:
\begin{equation}
\dot{v}(t)+\nu\, v(t)=0
\end{equation}
where
\begin{equation}
\nu=\int_0^\infty d\tau\,{\cal D}(\tau).
\label{18}
\end{equation}
Recalling that $\langle\dot{R}(t)\rangle=e^{i\Omega t}v(t)$, we may realize that the 
imaginary part of $\nu$ yields the shift of the cyclotron frequency
previously mentioned  in Eq. (\ref{1.4}), i.e.,
$\Omega_{\rm eff}=\Omega-{\rm Im}\,\nu$, while the real part, which
must be nonnegative, corresponds to the damping frequency $\nu_d$ defined in the same equation. 
Then, the transverse and
longitudinal friction coefficients arise from Eqs. (\ref{1.4p1}) and (\ref{1.4p2}) as:
\begin{eqnarray}
D'_M &=& (\rho_s\kappa/\Omega){\rm Im}\,\nu,\\
D_M &=&(\rho_s\kappa/\Omega){\rm Re}\,\nu,
\end{eqnarray}
 where the subscript $M$ indicates Markov approximation.
The real and imaginary parts of the frequency $\nu$, when considered as functions of $\Omega$,
obey Kramers-Kr\"onig relations\cite{ku} which lead to the following expression for the 
transverse friction coefficient:
\begin{equation}
D'_M(\Omega)=\frac{1}{\pi\Omega}\,{\rm P}\int_{-\infty}^\infty d\omega\frac{\omega D_M(\omega)}
{\omega-\Omega},
\label{DPP}
\end{equation}
where P denotes the Cauchy principal part and, 
\begin{eqnarray}
D_M(\omega) & = &
\frac{2\pi}{L\hbar\omega}\sum_{ {\bf k}, {\bf q}}  \, \, \delta_{k_zq_z}
({\bf k}-{\bf q})^2[|\Lambda(k,q)|^2(n_q-n_k)
\delta(\omega_k-\omega_q-\omega)\nonumber\\
&&+2|\Gamma(k,q)|^2(f_q-f_k)
\delta(\epsilon_k/\hbar-\epsilon_q/\hbar-\omega)].
\label{DOm}
\end{eqnarray}
The above even function of $\omega$ when evaluated at $\omega=\Omega$ gives the longitudinal
friction coefficient, and
it is easy to verify that $D_M(\Omega)>0$ since $\omega_k>\omega_q
\Rightarrow n_q>n_k$, and the same for the terms containing the fermion occupation numbers.
Note that only the scattering events that conserve energy will contribute to the longitudinal
friction coefficient (see the arguments of the Dirac deltas in (\ref{DOm}) for 
$\omega=\Omega$). This consequence of the Markov approximation can be physically understood 
in terms of the time-energy uncertainty principle. In fact, in the long time limit only the
microscopic states with the longest lifetimes are expected to remain with a nonnegligible
probability of undergoing a scattering transition, and according to the time-energy 
uncertainty principle, energy should be practically conserved at the end of such 
transitions.\cite{note}

We have studied in $I$ the phonon-roton contribution to the longitudinal friction
coefficient which arises from the first term inside the square brackets in (\ref{DOm}).
We showed that the limit of elastic scattering $\Omega\rightarrow 0$ yields an excellent
agreement with the values derived from experimental data for the roton temperature range,
provided the scattering amplitude $\Lambda$ is set to get agreement with the Iordanskii
results for the low-temperature phonon dominated regime. 
Let us now examine the $^3$He contribution to the
longitudinal friction in Eq. (\ref{DOm}) in the limit $\Omega\rightarrow 0$. Denoting such
a term by $D_3(\Omega)$, we have
\begin{eqnarray}
D_3(0)&=&\lim_{\Omega\rightarrow 0}D_3(\Omega)=-\frac{4\pi\hbar}{L}\sum_{ {\bf k} , {\bf q}}
 \, \, \delta_{k_zq_z}({\bf k}-{\bf q})^2\,|\Gamma(k,q)|^2\,
\frac{\partial f_k}{\partial\epsilon_k}\,\delta(\epsilon_k-\epsilon_q)\nonumber\\
&=& 
-\frac{2A^2\hbar}{(2\pi)^4}
\int d^3{\bf k}\,\int d^3{\bf q}\,
\,\frac{\partial f_k}{\partial\epsilon_k}
\,\,\delta(\epsilon_k-\epsilon_q)\,\,({\bf k}-{\bf q})^2\,\,
 \delta(k_z-q_z)\,
|\Gamma(q,k)|^2,
\label{3.7}
\end{eqnarray}
where the continuum limit was explicitly considered in the last expression, $A$ being the 
area of the system in the $x-y$ plane. Here most of the integration can be analytically 
performed in spherical coordinates, leading to the following one-dimensional integral:
\begin{equation}
D_3(0)=-\frac{4A^2m^*}{3\pi^2\hbar}\int_0^\infty dk\,k^4
|\Gamma(k,k)|^2\,
\frac{\partial f_k}{\partial\epsilon_k},
\label{3.8}
\end{equation}
where a Landau-Pomeranchuk dispersion relation $\epsilon_k=\hbar^2k^2/2m^*$ was utilized.
The above expression can be shown to be equivalent to the
R-R formula,\cite{reif}
provided the scattering amplitude fulfills
\begin{equation}
|\Gamma(k,k)|^2=\frac{9\pi\hbar^3}{128m^*A^2}u(k)\sigma(k),
\label{3.9}
\end{equation}
where $u(k)=\hbar k/m^*$ denotes the group velocity of $^3$He quasiparticles and $\sigma(k)$
 corresponds
to a total momentum-transfer cross section for vortex-$^3$He scattering, which, in a 
low $^3$He concentration regime,\cite{reif} can be approximated by the constant value
$\sigma_0=(18.3\pm 0.7)$ \AA.

\section{breakdown of the markov approximation. non-markovian treatment}
We have shown in $I$, that finite values of the cyclotron frequency extracted from
recent theories, yield values of the
 longitudinal friction coefficient of the order of that obtained
in the elastic limit $\Omega\rightarrow 0$. We shall henceforth work under such an assumption,
i.e., $D_M(\Omega)\sim D_M(0)$. Thus it can be shown that Eq. (\ref{DPP}) can be approximated
as follows:
\begin{equation}
D'_M(\Omega)\simeq\frac{2}{\pi\Omega}\int_{0}^\infty d\omega D_M(\omega),
\label{DP}
\end{equation}
where we have also assumed $D'_M(\Omega)>>D_M(\Omega)$.
The quasiparticle frequency cutoff in Eq. (\ref{DOm}) (roughly two times the roton frequency)
 yields
the same cutoff to the frequency $\omega$ in Eq. (\ref{DP}). This shows that the integral 
in (\ref{DP}) possesses a finite value and thus $D'_M(\Omega)$ would diverge
as $\Omega^{-1}$ in the 
limit $\Omega\rightarrow 0$. Later we shall see that this unphysical result arises from a
breakdown of the Markov approximation. In fact, one could expect the effective frequency 
$\Omega_{\rm eff}$ to be lower than the cyclotron frequency
by the effect of friction, but the existence of a critical cyclotron
frequency below which the effective frequency becomes negative, seems to be quite unphysical,
i.e., one would expect
effective frequency values bounded as $0<\Omega_{\rm eff}=
\Omega-\Omega D'_M(\Omega)/\rho_s\kappa<\Omega$.
We shall see in the following that a non-Markovian treatment yields in fact such bounds.
To see this, let us return to the integrodifferential equation (\ref{noM}) and take its
Laplace transform according to the definition $\tilde{v}(z)=\int_0^\infty \exp(izt)
v(t)dt$ (Im$z>0$):
\begin{equation}
\tilde{\cal D}(z) \tilde{v}(z)=iz\tilde{v}(z)+v(0),
\label{27}
\end{equation}
where the Laplace transform of the memory kernel (\ref{16}) reads as,
\begin{equation}
\tilde{\cal D}(z)=\frac{i(\Omega-z)}{\rho_s\kappa\Omega\pi}\int_{-\infty}^\infty
d\omega\frac{\omega D_M(\omega)}{\omega+\Omega-z}.
\label{28}
\end{equation}
Then from Eq. (\ref{27}) we have,
\begin{equation}
\tilde{v}(z)=\frac{v(0)}{-iz+\tilde{\cal D}(z)}
\label{29}
\end{equation}
and $v(t)$ arises from the singularities of $\tilde{v}(z)$ in the lower half-plane, Im$z<0$.
For instance, if the expression (\ref{29}) has a unique simple pole located at
$z_0=-i\tilde{\cal D}(z_0)$, we get
\begin{equation}
v(t)=v(0)e^{-iz_0t},
\end{equation}
and the Markov approximation would be valid provided $\tilde{\cal D}(z_0)\simeq
\tilde{\cal D}(0)$ (cf. Eq. (\ref{18})). 
That is, taking the limit $z\rightarrow i0^+$ in the Cauchy
integral of Eq. (\ref{28}) we get,\cite{bal}
\begin{equation}
\tilde{\cal D}(0)=\frac{i}{\rho_s\kappa\pi} {\rm P} \int_{-\infty}^\infty
d\omega\frac{\omega D_M(\omega)}{\omega+\Omega}\,+\,\frac{\Omega D_M(\Omega)}{\rho_s\kappa}
=\nu.
\end{equation}
Therefore, from Eq. (\ref{28}) we may realize that for such an approximation to be valid,
it necessarily should be $|z_0|=|\nu|<<\Omega$, i.e., $\Omega D_M(\Omega)/\rho_s\kappa<<\Omega$
and $\Omega D'_M(\Omega)/\rho_s\kappa<<\Omega$.
Now, according to the low-cyclotron frequency approximation (\ref{DP}), the last condition
will not be fulfilled for low enough frequencies, that is, the real part of $z_0$ will 
remain finite for $\Omega\rightarrow 0$. This suggests the following approximation
 to find the poles
of Eq. (\ref{29}):
\begin{equation}
iz_0=\tilde{\cal D}(z_0)\simeq \tilde{\cal D}({\rm Re}z_0)=
\frac{i(\Omega -{\rm Re}z_0) }{\rho_s\kappa\pi\Omega}\, {\rm P} \int_{-\infty}^\infty
\frac{\omega D_M(\omega)d\omega}{\omega+\Omega-{\rm Re}z_0}\,+\,
\frac{(\Omega-{\rm Re}z_0)^2  }{\rho_s\kappa\Omega} D_M(\Omega-{\rm Re}z_0),
\label{32}
\end{equation}
or, equivalently,
\begin{eqnarray}
{\rm Im}z_0&=&-\frac{(\Omega-{\rm Re}z_0)^2  }{\rho_s\kappa\Omega} D_M(\Omega-{\rm Re}z_0)\\
{\rm Re}z_0&=&
\frac{(\Omega -{\rm Re}z_0) }{\rho_s\kappa\pi\Omega}\, {\rm P} \int_{-\infty}^\infty
\frac{\omega D_M(\omega)d\omega}{\omega+\Omega-{\rm Re}z_0}\simeq
\frac{2(\Omega -{\rm Re}z_0) }{\rho_s\kappa\pi\Omega} \int_0^\infty
D_M(\omega)d\omega,
\label{34}
\end{eqnarray}
where the last equality arises from the approximation (\ref{DP}), i.e., assuming
$D_M(0)\sim D_M(\Omega_{\rm eff})<<\int_0^\infty d\omega D_M(\omega)/\Omega_{\rm eff}$,
($\Omega_{\rm eff}=\Omega-{\rm Re}z_0$).
Then from Eq. (\ref{34}) we get the solution
\begin{equation}
\Omega_{\rm eff}=\Omega-{\rm Re}z_0=\Omega/\{1+[2/(\rho_s\kappa\pi\Omega)]\int_0^\infty 
D_M(\omega)d\omega\},
\label{35}
\end{equation}
where, in fact, the effective frequency turns out to be bounded according to our previous
discussion, $0<\Omega_{\rm eff}<\Omega$. Note also that $\Omega_{\rm eff}
\rightarrow 0$ corresponds to the limit of a vanishing cyclotron frequency, as expected.
Finally, the friction coefficients $D=-(\rho_s\kappa/\Omega){\rm Im}z_0$ and
$D'=(\rho_s\kappa/\Omega){\rm Re}z_0$ reads as,
\begin{eqnarray}
D &=& (\Omega_{\rm eff}/\Omega)^2 D_M(\Omega_{\rm eff})\label{36}\\
D' &=& \rho_s\kappa(1-\Omega_{\rm eff}/\Omega),
\label{37}
\end{eqnarray}
which generalize the previous Markovian expressions (\ref{DOm}) and (\ref{DP}).
Here it is expedient to recall that (\ref{36}) and (\ref{37}) were extracted under the 
approximations of Eqs. (\ref{DP}) and (\ref{32}), both being equivalent to 
$D<<D'$. If, in addition, we have
$D'<<\rho_s\kappa$, then $\Omega_{\rm eff}\simeq\Omega$ and Eqs. (\ref{36}) and (\ref{37}) tend
 to
the Markovian expressions. In other words, the frequency ratio $\Omega_{\rm eff}/\Omega$
can be thought of as a measure of the proximity to the Markovian limit.
Note that the limit of a vanishing cyclotron frequency is a strongly non-Markovian one,
with $D\sim {\cal O}(\Omega^2)$ 
and $\rho_s\kappa-D'\sim {\cal O}(\Omega)$. Such a behavior of the transverse
coefficient corresponds to the lower limit of $\Omega_{\rm eff}\sim {\cal O}(\Omega^2)$ as 
given
 by 
Eq. (\ref{35}). As regards the longitudinal coefficient, since the effective frequency was 
absent from our previous analysis in $I$, the limiting values for elastic 
scattering ($\Omega\rightarrow 0$) there reported are now drastically changed to vanishing
values. 
However, we shall next see that
there is a range of cyclotron frequency values that keep
the previous agreement with the experimental values up to temperatures near 1 K.

\section{analysis of results for a pure $^4$h\lowercase{e} system}
\subsection{Longitudinal friction coefficient and the cyclotron frequency range}
\label{secVA}
In Table I we may compare values of the longitudinal friction
coefficient 
computed from Eq. (\ref{36}) for $\Omega$=0.01$-$0.03
ps$^{-1}$, with the corresponding values arising from the Iordanskii formula\cite{ior}
(phonon range), plus the R-R formula\cite{reif,dijk}
(roton range).
\begin{table}
\caption{Longitudinal friction coefficient [10$^{-6}$g cm$^{-1}$ s$^{-1}$] versus temperature
for a pure $^4$He system.
The values in the third and fifth columns were calculated from Eq. (\ref{36}) and have to
be compared with the corresponding values in the second column, which arise from 
Refs.~\onlinecite{reif},~\onlinecite{ior} and~\onlinecite{dijk}.
The values in the third and fourth columns were calculated for $\Omega=0.01$ ps$^{-1}$,
while the values in the fifth and sixth ones correspond to $\Omega=0.03$ ps$^{-1}$.
Powers of 10 are enclosed in brackets.}
\begin{ruledtabular}
\begin{tabular}{cccccc} 
$T[{\rm K}]$ 
& $D_{\rm Refs}$  &
 $D_{0.01}$ & $\Omega_{\rm eff}/\Omega$ &
 $D_{0.03}$ & $\Omega_{\rm eff}/\Omega$  \\
\hline
0.1 & 2.61[-8] & 2.63[-8] & 1.000 & 2.97[-8] & 1.000 \\
0.2 & 8.34[-7] & 8.15[-7] & 0.999 & 8.45[-7] & 1.000 \\
0.3 & 6.33[-6] & 5.98[-6] & 0.996 & 6.11[-6] & 0.999 \\
0.4 & 2.84[-5] & 2.59[-5] & 0.989 & 2.63[-5] & 0.996 \\
0.5 & 2.06[-4] & 2.07[-4] & 0.979 & 1.95[-4] & 0.993 \\
0.6 & 2.43[-3] & 2.53[-3] & 0.962 & 2.39[-3] & 0.987 \\
0.7 & 1.79[-2] & 1.80[-2] & 0.938 & 1.77[-2] & 0.979 \\
0.8 & 8.27[-2] & 7.67[-2] & 0.900 & 8.07[-2] & 0.964 \\
0.9 & 0.274 & 0.221 & 0.840 & 0.256 & 0.940\\
1.0 & 0.714 & 0.460 & 0.749 & 0.618 & 0.900\\
\end{tabular}
\end{ruledtabular}
\end{table}
Such calculations were performed taking into account only the phonon-roton 
contribution in Eq. (\ref{DOm}), i.e., for a pure $^4$He system.
In the phonon dominated regime ($T<0.4$ K), the lack of experimental data leads us to 
compare with the results stemming from the Iordanskii formula, i.e., those 
arising from the 
 limit $\Omega\rightarrow 0$ of Eq. (\ref{DOm}).\cite{ca1}
Actually, it will be enough to have $\hbar\Omega/k_BT<1$ in order to keep $D_M(\Omega)\simeq
D_M(0)$ (cf. the figures for $T=0.1$ K in Table I), but too small values of $\Omega$ would
affect the Markovian approximation $\Omega_{\rm eff}\simeq\Omega$,
leading to appreciable discrepancies between 
the Iordanskii result, $D_M(0)$, and our expression (\ref{36}).
The phonon influence becomes negligible for $T\gtrsim
0.65$ K, allowing the comparison of our results
with those arising from the R-R formula,\cite{reif} which constitutes a good fit
to experimentally derived values up to temperatures about 1.3 K.\cite{dijk}
Taking into account
an estimated uncertainty of order 10 \%, we may see from Table I that our
results lie within such error bounds for temperatures up to 0.8 K (0.9 K)  for $\Omega=
0.01$ ps$^{-1}$ (0.03 ps$^{-1}$). Note that for $T$=1 K, our result for 
$\Omega=
0.01$ ps$^{-1}$ (0.03 ps$^{-1}$) falls 30 \% (4 \%) below the lower error boundary.
To summarize, we have identified a narrow range of cyclotron frequency values 
(0.01 ps$^{-1}\lesssim\Omega\lesssim$0.03 ps$^{-1}$) yielding longitudinal friction
figures which are in agreement with the Iordanskii formula and experimental data.
However, it is important to note that according to recent theories,\cite{duan,arovas,tang}
which show that the cyclotron frequency scales down {\it logarithmically} with the vortex
size, such a range of $\Omega$ would be consistent with a relatively wide range of 
macroscopic sizes of the system.

\subsection{\label{VB}
Effective frequency, memory effects and transverse friction coefficient}
Another feature of Table I shows that the frequency ratio $\Omega_{\rm eff}/\Omega$ decreases
with increasing temperature, which reflects a corresponding increase of $D'$ (Eq. 
(\ref{37})). Particularly, at the lowest temperatures, the Markov approximation
$\Omega_{\rm eff}/\Omega\simeq 1$ shows to be excellent, becoming gradually less adequate as
the temperature increases. The highest temperature range (0.9$-$1.0 K) displays the largest
differences with the Markov approximation, as well as the highest discrepancy with the
experimental results. In other words, the longitudinal friction phenomenon appears to be
consistent with weakly, at most, non-Markovian processes ($\Omega_{\rm eff}/\Omega >0.9$).
There is, however, another possible interpretation of such a discrepancy with the 
experimental results for $\Omega_{\rm eff}/\Omega <0.9$, which is related to an eventual
failure of the weak-coupling approximation. In fact, according to Eqs. (\ref{35}) and
(\ref{DOm}), we may see that for fixed $\Omega$, the parameter $\Omega_{\rm eff}/\Omega$
will behave as a decreasing function of the coupling strengths $\Lambda$ and $\Gamma$,
such that $\Omega_{\rm eff}/\Omega\rightarrow 1$ for a vanishing coupling ($D_M(\omega)
\rightarrow 0$), while $\Omega_{\rm eff}/\Omega\rightarrow 0$ for an infinite coupling
($D_M(\omega)\rightarrow \infty$). Then, only a higher portion of the interval  
$0<\Omega_{\rm eff}/\Omega <1$ should be expected to be consistent with a weak-coupling
approximation. In conclusion, the above discrepancy with the experimental results for
$\Omega_{\rm eff}/\Omega <0.9$ may also be regarded as an indication of a possible failure
of the weak-coupling approximation.

The transverse friction coefficient, in contrast to the longitudinal one, possesses a 
strong dependence on $\Omega$. In fact, Eq. (\ref{DP}) shows that in the Markovian limit,
the bounds 0.01 ps$^{-1}<\Omega<$0.03 ps$^{-1}$ lead to a factor  3 of spreading 
[$D'_M$(0.01 ps$^{-1}$)=3$D'_M$(0.03 ps$^{-1}$)], while the non-Markovian figures of Table I
can reduce such a factor somewhat ($>$2.5). Another important difference between both 
friction coefficients stems from the degree of dependence on the quasiparticle dispersion 
relation
cutoff features.\cite{hsu,pistolesi}
 On the one hand, the longitudinal coefficient, which depends mainly on
$D_M(\Omega_{\rm eff})$, turns out to be almost independent of such a cutoff, since 
$\Omega_{\rm eff}$ and $\Omega$ are two orders of magnitude lower than the roton frequency.
The transverse coefficient, on the other hand, being mainly dependent on the integral
$\int_0^\infty D_M(\omega)d\omega$, has therefore an important dependence on the quasiparticle
 cutoff
through the corresponding dependence of the scattering amplitudes $\Lambda$ and $\Gamma$,
which is mostly uncertain. In summary, due to the above uncertainties in the calculation of
the transverse coefficient, only its order of magnitude should be reliable which, nevertheless,
turns out to be quite useful to ensure that the condition $D<<D'$ is fulfilled.
Recall that such a condition was assumed in the derivation of Eqs. (\ref{36}) and (\ref{37}),
and in fact, taking into account that 
$\rho_s\kappa\simeq 145\times 10^{-6}$ g cm$^{-1}$s$^{-1}$,
all the values of Table I can be shown to be consistent with $D<<D'$.
It is worthwhile recalling also the lack of experimental results on $D'$ for temperatures 
below 1.3 K. Taking into account only the vortex drag due to the scattering of rotons,
the transverse coefficient can be written in terms of a transverse scattering
length $\sigma_\perp$, viz. $D'=\rho_n v_G\sigma_\perp$, where $\rho_n$ denotes the normal
fluid density and $v_G$ the average group velocity of thermal rotons.\cite{bar,don} 
However, only speculative
assumptions about the form of $\sigma_\perp$ for temperatures below 1.3 K were
reported.\cite{bar}
 In addition, it has been argued that the so-called Iordanskii force\cite{ior}
gives rise to an additional transverse coefficient to be substracted from $D'$, yielding
a total transverse coefficient\cite{bar,don} given by $D_t=D'-\rho_n\kappa$.
 However, the sign, amplitude, and existence of this Iordanskii force are
still subject to debate.\cite{sonin,wex,fort,thou1} Recently, Fortin\cite{fort} has applied the
formalism of Thouless, Ao, and Niu\cite{thou} to compute the transverse and longitudinal
coefficients due to the scattering of noninteracting phonons in two dimensions. He finds
a transverse coefficient which turns out to be of opposite sign to ours and to that of Refs.
\onlinecite{bar} and~\onlinecite{don},
 which is interpreted in terms of a negative vortex mass due
to phonons. Such a discrepancy in the sign stems from the fact that, according to his 
equations, the transverse and longitudinal coefficients would be related, as functions of the
cyclotron frequency, by Kramers-Kr\"onig relations, while in our case such relations are 
connecting instead
the real and imaginary parts of the Markovian frequency $\nu$ (cf. Eq. (\ref
{DPP})).

\subsection{\label{VC}Phonon emission and memory effects at zero temperature}
At this point, as a useful
 complement to our study, it will be instructive to discuss in some
detail the memory effects related to phonon emission at zero temperature. 
We will base our analysis on the theory developed by Arovas and Freire\cite{arovas}
for vortex dynamics in superfluid films.
In fact, suppose that 
the vortex is set in motion at positive times by the action of a homogeneous time dependent
superfluid flow, i.e., it is assumed that both, the superfluid velocity ${\bf v}_s$ 
and the vortex 
velocity ${\bf v}$,
 are zero for negative times. Then, the vortex equation of motion can be written
\cite{arovas}
\begin{equation}
\int_0^t M(\tau)\dot{V}(t-\tau)d\tau=i\rho_s\kappa[V(t)-V_s(t)].
\label{38}
\end{equation}
The right-hand side of this equation corresponds to the usual Magnus force $\rho_s\kappa
\hat{\bf z}
\times({\bf v}-{\bf v}_s)$ (cf. Eq. (\ref{1.1}))
expressed in complex notation ($V=v_x+iv_y$), while the 
left-hand side will differ from the familiar Newtonian product of mass times 
acceleration, unless the
memory or causal\cite{arovas} kernel $M(\tau)$ has a negligible lifetime. The memory, which
actually plays an important role in this case, stems from the vortex coupling to the low 
lying excitations of the superfluid (phonons), in close analogy to the retardation and 
radiation
effects stemming from electron-photon coupling in electrodynamics\cite{arovas}. Now, we focus
upon the long time limit of Eq. (\ref{38}). That is, for $t>>$lifetime of $M(\tau)$,
 the left-hand side could be approximated as $\tilde{M}\dot{V}(t)$, where the effective 
vortex
mass $\tilde{M}$ is given by the Fourier (Laplace) transform of the memory kernel $M(\tau)$
at zero frequency, $\tilde{M}=\int_0^\infty d\tau M(\tau)=M'+iM''$,
 the imaginary part $M''$
being related to the dissipation of vortex energy in the form of phonon emission.
Then the solution of Eq. (\ref{38}) in the long time limit could be
easily obtained for constant
$V_s$:
\begin{equation}
V(t)=V_s\{1-\exp[i(\Omega+i\nu_r)t]\},
\label{39}
\end{equation}
where the cyclotron $\Omega$ and radiation damping $\nu_r$
frequencies respectively reads  as,
\begin{eqnarray}
\label{40}
\Omega &=& \frac{\rho_s\kappa M'}{M'^2+M''^2}\\
\label{41}
\nu_r &=& \frac{-\rho_s\kappa M''}{M'^2+M''^2}.
\end{eqnarray}
$M'$, however, is shown to diverge for an unbounded two-dimensional
system\cite{arovas} ($M'(\omega
\rightarrow 0)\sim -\ln \omega$), while for a finite macroscopic system one should expect
a corresponding finite value of $M'(>>|M''|)$, leading to the familiar expression (\ref{1.3})
for the 
cyclotron frequency in (\ref{40}), with $M'=m_v$ the vortex mass per unit length. 
As regards the radiation damping frequency, from Eq. (14) of Ref.~\onlinecite{arovas} we have 
$|M''(\omega\rightarrow 0)|=\kappa^2\rho_s/(8c_s^2)$, ($c_s$ = sound velocity) and then,
$\nu_r=\kappa(\Omega/c_s)^2/8$. This result derives as well from the expression for 
the mean power radiated by unit length of a vortex performing cyclotron motion (see Eq. (2.11)
of Ref.~\onlinecite{vinen}). Therefore,
 the radiation damping should be weak $\nu_r\sim 10^{-3}\Omega$ for our range of cyclotron 
frequency values ($\Omega\sim 0.01$ ps$^{-1}$), whereas it would be relatively
strong, $\nu_r\sim\Omega$,
for cyclotron frequencies arising from a hydrodynamical model for the vortex mass,\cite{don}
$\Omega\approx 3$ ps$^{-1}$.
It is interesting to notice that a
 vanishing  damping frequency for an infinite system in (\ref{41})  does not
preclude the approach of the vortex velocity to the superfluid velocity at long times.
Actually, it simply means that such an approach will be slower than the exponential one
of Eq. (\ref{39}). To see this, let us 
integrate by parts
the left-hand side of Eq. (\ref{38}) getting $M(0)V(t)+
\int_0^t \dot{M}(\tau)V(t-\tau)d\tau$. 
Then, approximating in the long time limit the last 
integral  as $\int_0^t \dot{M}(\tau)V(t)d\tau$, the left-hand side of Eq. (\ref{38}) turns 
out to
be simply $M(t)V(t)$, from which we get
\begin{equation}
V(t)=V_s\left[\frac{1-iM(t)/\rho_s\kappa}{1+(M(t)/\rho_s\kappa)^2}\right],
\end{equation}
where\cite{arovas} $M(t)/\rho_s\kappa\simeq\xi/(2c_st)$, $\xi=\kappa/(2\pi c_s)$ being
the coherence length. Thus, the above expression shows that
in the case of an infinite system, the approach of the vortex 
velocity to the superfluid velocity
turns out to be, in contrast to (\ref{39}), a slow nonexponential one.

\section{analysis of results for dilute solutions of
$^3$h\lowercase{e} in $^4$h\lowercase{e} }
In case of a $^3$He-$^4$He mixture we have to take into account both terms in the expression
(\ref{DOm}) for $D_M(\omega)$. The calculation of the fermion term $D_3(\omega)$
(cf. Eq. (\ref{3.7})) turns out to be similar to that leading to $D_3(0)$ in Eq. (\ref{3.8}),
namely
\begin{equation}
D_3(\omega)=\frac{m^*A^2}{\pi^2\hbar^2\omega}\int_0^\infty dk |\Gamma(k,q)|^2\,(f_k-f_q)\,
k^2(q^2+k^2/3),
\label{6.1}
\end{equation}
where the value of the momentum $q$ arises from the argument of the second Dirac delta in 
Eq. (\ref{DOm}), i.e., $q^2=k^2+2m^*\omega/\hbar$. 
Thus, the cutoff of the Landau-Pomeranchuk dispersion relation imposes the same cutoff
($\simeq 1$ ps$^{-1}$, see Ref.~\onlinecite{hsu}) to the
frequency $\omega$ in Eq. (\ref{6.1}). Recall that the cutoff uncertainties
 will be reflected in the evaluation of the transverse
friction coefficient, as was already mentioned in Sec. \ref{VB}.
To compute Eq. (\ref{6.1}) we 
utilized the following simple generalization of the expression (\ref{3.9}) for $k\neq q$:
\begin{equation}
|\Gamma(k,q)|^2=\frac{9\pi}{128}\frac{\hbar^4}{(m^*A)^2}\,\sigma_0\sqrt{kq}.
\label{6.2}
\end{equation}

In Table II we may compare some experimental results for the longitudinal friction coefficient
(third column), with the corresponding results
arising from our approach, along with the values computed from the Iordanskii\cite{ior} and
R-R\cite{reif} formulas. 
\begin{table}
\caption{Longitudinal friction coefficient [10$^{-6}$g cm$^{-1}$ s$^{-1}$] versus temperature
and $^3$He concentration for mixtures. The third column corresponds to experimental results, 
while the forth one corresponds to values arising from Iordanskii formula\cite{ior}
(phonon contribution)
plus R-R formulas\cite{reif} (roton+$^3$He contribution). The remaining notation
is the same as in Table I.}
\begin{ruledtabular}
\begin{tabular}{cccccccccc} 
$T[{\rm K}]$ 
& C & $D_{\rm exp}$ & $D_{\rm Refs}$  &
 $D_{0.01}$ & $\Omega_{\rm eff}/\Omega$ & 
$D_{0.02}$ & $\Omega_{\rm eff}/\Omega$ &
 $D_{0.03}$ & $\Omega_{\rm eff}/\Omega$  \\
\hline
0.28 & 2.84[-5] & 4.69[-3] & 4.69[-3] & 4.69[-3] & 0.995 & 4.69[-3] & 0.997 &
4.64[-3] & 0.998 \\
0.28 & 7.55[-6] & 1.35[-3] & 1.25[-3] & 1.25[-3] & 0.996 & 1.25[-3] & 0.998 &
1.24[-3] & 0.999\\
0.61 & 7.55[-6] & 4.56[-3] & 5.01[-3] & 4.87[-3] & 0.960 & 4.90[-3] & 0.980 &
4.80[-3] & 0.986\\
0.615 & 1.4[-7] & 3.34[-3] & 3.43[-3] & 3.56[-3] & 0.959 & 3.52[-3] & 0.979 &
3.39[-3] & 0.986\\
0.643 & 1.4[-7] & 6.04[-3] & 6.17[-3] & 6.36[-3] & 0.953 & 6.32[-3] & 0.976 &
6.11[-3] & 0.984\\
0.67 & 1.4[-7] & 1.01[-2] & 1.04[-2] & 1.07[-2] & 0.947 & 1.07[-2] & 0.973 &
1.04[-2] & 0.982\\
\end{tabular}
\end{ruledtabular}
\end{table}
Given the low $^3$He concentrations of Table II, the 
Fermi temperatures turn out to be at most two orders of magnitude below the experimental
ones, and so the Fermi occupation numbers in Eqs. (\ref{3.8}) and (\ref{6.1}) can be
very well approximated by the Maxwell-Boltzmann statistics. At $T$=0.28 K the phonon 
contribution to the friction, stemming from the Iordanskii formula, turns out to be negligible
in comparison to the $^3$He term given by Eqs. (\ref{3.8}) and (\ref{3.9}) (cf. the values
of $D_{\rm Refs}$ in Tables I and II). Then, the experimental value $D_{\rm exp}$
for $T$=0.28 K and $C$=2.84$\times 10^{-5}$
 in Table II (being $C=n_3/(n_3+n_4)$, $n_i$= number density
of $^i$He atoms), which was measured within an error less than 1\%, was utilized by R-R
 to calculate the effective cross section $\sigma_0=18.3$ \AA $\,$
 in Eq.
(\ref{3.9}), assuming for the effective mass the value $m^*=2.5m_3$ ($m_3$= actual mass of
a $^3$He atom). A similar procedure was followed in our case, since the value of $\sigma_0$
in (\ref{6.2}) was set to get agreement with the experimental value $D_{\rm exp}$
=4.69$\times 10^{-3}$ for $\Omega=0.02$ ps$^{-1}$, i.e., the center of the cyclotron frequency
range discussed in Section \ref{secVA}. Then, equating (\ref{36}) to 4.69$\times 10^{-3}$,
we extracted 
the value $\sigma_0=18.54$ \AA, which turns out to be slightly greater than the
R-R result. Although the phonon contribution to $D_M$ in Eq. (\ref{36}) is in fact
negligible for $T$=0.28 K and $C$=2.84$\times 10^{-5}$,
 this is not the case for $\Omega_{\rm eff}/
\Omega$ (Eq. (\ref{35})), since the phonon contribution to the integral $\int_0^\infty
D_M(\omega)d\omega$ turns out to be greater than the $^3$He one. Nevertheless, as seen from
Table II, the factor $(\Omega_{\rm eff}/\Omega)^2$ in (\ref{36}) remains close to unity and
then, in practice, almost all the longitudinal friction should be ascribed to $^3$He 
scattering.

In a second experiment,
to prove the proportionality of the friction coefficient to the $^3$He number density in the
dilute limit, R-R performed a measure for the same temperature $T$=0.28 K and
a smaller concentration of $C=7.55\times 10^{-6}$. They obtained the value $D_{\rm exp}=
(1.35\pm 0.06)\times 10^{-3}$, which turns out to be in agreement with their theoretical
calculation arising from (\ref{3.8}), within the limits of estimated error.
As regards our calculation from Eq. (\ref{36}), it yields practically the same figures as
the R-R formula (Table II). 

The diluted sample was also used to verify the 
additivity of $^3$He and roton scattering, by performing an experiment at the relatively high
temperature of 0.61 K. The directly measured value $D_{\rm exp}=4.56\times 10^{-3}$ was then
contrasted with that arising from the addition of R-R formulas for $^3$He  and roton
scattering contributions, namely 4.79$\times$10$^{-3}$. The phonon
contribution, on the other hand,
 was ignored, presumably because of some discrepancies arisen from
the Pitaevskii's\cite{pita}
calculation of the phonon-scattering cross section. Actually, taking into
account such a contribution, the value of the friction coefficient should have been increased 
to 
5.15$\times$10$^{-3}$. The Pitaevskii's result was later modified by Iordanskii
\cite{ior} in that the coefficient of proportionality to $T^5$ of the friction coefficient
due to phonons
 was shown to be smaller by a factor $\sim 0.62$. That is, taking into account the phonon 
contribution stemming from the Iordanskii formula, the corrected value 5.01$\times$10$^{-3}$
(Table II) is in fact closer to the experimental one.
Finally, we compare with our results computed from Eq. (\ref{36}). From Table II we see that 
such results are closer to the observed one than the previous estimation of Iordanskii+R-R, and 
 in this better agreement it is 
important to remark the role played by the memory effect, 
which is embodied in the factor 
$(\Omega_{\rm eff}/\Omega)^2<1$ in Eq. (\ref{36}).

Next we analyze a set of measures performed for ordinary helium ($C=1.4
\times 10^{-7}$) at temperatures of 0.615, 0.643 and 0.67 K.
The second measure ($T=0.643$ K) was reported in Ref.~\onlinecite{rayfield},
 while the remaining
two ones are included in Ref.~\onlinecite{reif}. Under such conditions,
the incidence
of $^3$He scattering is almost negligible in both, the Iordanskii+R-R results and our
figures computed from (\ref{36}). 
From Table II we see that, analogously to the above results for
$T$=0.61 K, theoretical calculations
again overestimate the experimental data, and the best
agreement is also obtained for our result at $\Omega=0.03\,{\rm ps}^{-1}$. 

Finally, from Table II we notice that the memory effect remains small ($\Omega_{\rm eff}/
\Omega\gtrsim 0.95$), showing the same increase with temperature as in Table I.
On the other hand, the dependence of $\Omega_{\rm eff}/\Omega$ on concentration is not 
clear from Table II except for $T$=0.28 K, where we find a slight reduction for a higher
concentration. It is not difficult, however, to generalize such a result taking into 
account that in the dilute limit, the Maxwell-Boltzmann approximation to $f_k$  and $f_q$
in Eq. (\ref{6.1}) yield a $D_3(\omega)$ proportional to the $^3$He number density, which
in turn implies a growing of $\int_0^\infty D_M(\omega)d\omega$ with concentration and,
accordingly, a decreasing behavior for $\Omega_{\rm eff}/\Omega$ in Eq. (\ref{35}).
In other words, $\Omega_{\rm eff}/\Omega$ is shown to be a decreasing function of the 
number of quasiparticles, i.e., phonons, rotons and $^3$He impurities, interacting with
the vortex.

We hope, in concluding this report,
 that the present theoretical results will stimulate an experimental
investigation of degenerate-$^3$He richer vortex friction regimes.

\appendix*
\section{derivation of the vortex equation of motion}
Our starting point is the Hamilton
equation of motion for the creation operator of right circular quanta, whose time derivative
turns out to be proportional to the vortex velocity:
\begin{equation}
\dot{a}^\dagger=\sqrt{\pi\rho_sL/m_4}\dot{R},
\end{equation}
where $L$ denotes the vortex line length. Thus we
have,
\begin{equation}
\dot{a}^\dagger=\frac{i}{\hbar}[H,a^\dagger]=i\Omega a^\dagger+\hat{O},
\end{equation}
where $\hat{O}$ denotes the quasiparticle operator:
\begin{equation}
\hat{O}=\frac{i}{\hbar\sqrt{\pi\rho_sL/m_4}}
\sum_{ {\bf k} , {\bf q},\sigma }  \, \, \delta_{k_zq_z}
[(k_y-q_y)+i(q_x-k_x)]
[\Lambda(k,q) b_{{\bf k}}^\dagger \,  b_{ {\bf q}}+
\Gamma(k,q) c_{{\bf k},\sigma}^\dagger \,  c_{ {\bf q},\sigma}].
\label{O}
\end{equation}
Next we consider the Hamilton equation for
$\dot{a}^\dagger$:
\begin{eqnarray}
\ddot{a}^\dagger & = & \frac{i}{\hbar}[H,\dot{a}^\dagger]=i\Omega \dot{a}^\dagger+
\frac{i}{\hbar}[H,\hat{O}]\nonumber\\
&=& i\Omega \dot{a}^\dagger -
\frac{1}{\hbar\sqrt{\pi\rho_sL/m_4}}
\sum_{ {\bf k} , {\bf q},\sigma }  \, \, \delta_{k_zq_z}
[(k_y-q_y)+i(q_x-k_x)]
[\Lambda(k,q)(\omega_k-\omega_q) b_{{\bf k}}^\dagger \,  b_{ {\bf q}}\nonumber\\
&&+
\frac{1}{\hbar}\Gamma(k,q) (\epsilon_k-\epsilon_q)
c_{{\bf k},\sigma}^\dagger \,  c_{ {\bf q},\sigma}]-
\frac{2i}{\hbar^2\pi\rho_sL/m_4}
\sum_{ {\bf k'}, {\bf k}} \sum_{ {\bf q},\sigma }  \, \, \delta_{k_zk'_z} \delta_{k'_zq_z}
[(q_x-k'_x)(k'_y-k_y)\nonumber\\
&&+(k'_y-q_y)(k'_x-k_x)] 
[\Lambda(k,k')\Lambda(k',q) b_{{\bf k}}^\dagger \,  b_{ {\bf q}}
+\Gamma(k,k')\Gamma(k',q)
c_{{\bf k},\sigma}^\dagger \,  c_{ {\bf q},\sigma}]a^\dagger.
\label{ddota}
\end{eqnarray}
As in $I$, we shall make use of a weak-coupling approximation, which consists in
retaining only second order terms in the scattering amplitudes $\Lambda$ and $\Gamma$.
This approximation is discussed in Sec. \ref{VB}. Then to 
approximate the above expression, we note that
to the zeroth order in such parameters we have
\begin{eqnarray}
a^\dagger & = & -i\,\dot{a}^\dagger/\Omega\\
b_{{\bf k}}^\dagger(t) \,  b_{ {\bf q}}(t) & = & e^{i(\omega_k-\omega_q)t}\,
b_{{\bf k}}^\dagger(0) \,  b_{ {\bf q}}(0)\\
c_{{\bf k},\sigma}^\dagger(t) \,  c_{ {\bf q},\sigma}(t) & = & e^{i(\epsilon_k-
\epsilon_q)t/\hbar}
c_{{\bf k},\sigma}^\dagger(0) \,  c_{ {\bf q},\sigma}(0),
\end{eqnarray}
while the first order arises from the Hamilton 
equations for $b_{{\bf k}}^\dagger \,  b_{ {\bf q}}$ and 
$c_{{\bf k},\sigma}^\dagger \,  c_{ {\bf q},\sigma}$. We have, for instance,
\begin{eqnarray}
\frac{d}{dt}(b_{{\bf k}}^\dagger \,  b_{ {\bf q}}) & = & \frac{i}{\hbar}[H,
b_{{\bf k}}^\dagger \,  b_{ {\bf q}}]\nonumber\\
& = & i(\omega_k-\omega_q) b_{{\bf k}}^\dagger \,  b_{ {\bf q}}
+\frac{i}{\hbar\sqrt{\pi\rho_sL/m_4}}
\{\sum_{ {\bf k'}}  \, \, \delta_{k'_zk_z}\Lambda(k',k)
[[(k_y-k'_y)+i(k_x-k'_x)]a^\dagger\nonumber\\
&& +[(k'_y-k_y)+i(k_x-k'_x)]a]\,b_{{\bf k'}}^\dagger \,  b_{ {\bf q}}
-\sum_{ {\bf q'}}  \, \, \delta_{q'_zq_z}\Lambda(q,q')
[[(q'_y-q_y)+i(q'_x-q_x)]a^\dagger\nonumber\\
&&+[(q_y-q'_y)+i(q'_x-q_x)]a]\,b_{{\bf k}}^\dagger \,  b_{ {\bf q'}}\},
\end{eqnarray}
and we may find a formal solution to this equation in 
$b_{{\bf k}}^\dagger(t) \,  b_{ {\bf q}}(t)$ by noting that the only dependence on
$b_{{\bf k}}^\dagger \,  b_{ {\bf q}}$ on the right-hand side comes from the first term.
Thus we have to the first order in $\Lambda$:
\begin{eqnarray}
b_{{\bf k}}^\dagger(t) \,  b_{ {\bf q}}(t)&=& e^{i(\omega_k-\omega_q)t}\,
b_{{\bf k}}^\dagger(0) \,  b_{ {\bf q}}(0)+\int_0^t d\tau \,
e^{i(\omega_k-\omega_q)\tau}
\frac{1}{\hbar\sqrt{\pi\rho_sL/m_4}}
\{\sum_{ {\bf k'}}  \, \, \delta_{k'_zk_z}\Lambda(k',k)
[[(k_y-k'_y)\nonumber\\
&&+i(k_x-k'_x)]\dot{a}^\dagger(t-\tau)/\Omega
+[(k_y-k'_y)+i(k'_x-k_x)]\dot{a}(t-\tau)/\Omega]\,e^{i(\omega_{k'}-\omega_q)(t-\tau)}
b_{{\bf k'}}^\dagger(0) \,  b_{ {\bf q}}(0)\nonumber\\
&&-\sum_{ {\bf q'}}  \, \, \delta_{q'_zq_z}\Lambda(q,q')
[[(q'_y-q_y)+i(q'_x-q_x)]\dot{a}^\dagger(t-\tau)/\Omega+[(q'_y-q_y)\nonumber\\
&&+i(q_x-q'_x)]\dot{a}(t-\tau)
/\Omega]\,e^{i(\omega_k-\omega_{q'})(t-\tau)}
b_{{\bf k}}^\dagger(0) \,  b_{ {\bf q'}}(0)\},
\label{b+b}
\end{eqnarray}
and analogously we may find a similar expression for 
$c_{{\bf k},\sigma}^\dagger(t) \,  c_{ {\bf q},\sigma}(t)$. 
Finally, replacing Eq. (\ref{b+b}) (and the corresponding expression for 
$c_{{\bf k},\sigma}^\dagger(t) \,  c_{ {\bf q},\sigma}(t)$) in Eq. (\ref{ddota}) and taking
mean values according to $\langle b_{{\bf k}}^\dagger(0) \,  b_{ {\bf q}}(0)\rangle = 
\delta_{{\bf k}{\bf q}}n_k$ and 
$\langle c_{{\bf k,\sigma}}^\dagger(0) \,  c_{ {\bf q},\sigma}(0)\rangle=
\delta_{{\bf k}{\bf q}}f_k$, we obtain Eq. 
(\ref{noM}).


\end{document}